\documentclass[twocolumn]{bmcart}

\usepackage[utf8]{inputenc} 
\usepackage{amssymb}
\usepackage[pdftex]{graphicx}
\usepackage{epstopdf}
\usepackage[fleqn]{amsmath}
\graphicspath{{bmc_template/}}

\startlocaldefs
\endlocaldefs

\begin{document}
	
	\begin{frontmatter}
		
		\begin{fmbox}
			\dochead{Research}
			
			
			\title{VLSI Architecture of Compact Non-RLL Beacon-based Visible Light Communication Transmitter and Receiver}
			
			
			\author[
			addressref={aff1},                   
			corref={aff1},                       
			email={nguyen.phuc.ni6@is.naist.jp}   
			]{\inits{DPN}\fnm{Duc-Phuc} \snm{Nguyen}}
			\author[
			addressref={aff1},
			email={le.dung.ku9@is.naist.jp}
			]{\inits{DDL}\fnm{Dinh-Dung} \snm{Le}}
			\author[
			addressref={aff1},
			email={hong@is.naist.jp}
			]{\inits{THT}\fnm{Thi-Hong} \snm{Tran}}
			\author[
			addressref={aff2},
			email={hhthuan@hcmus.edu.vn}
			]{\inits{HTH}\fnm{Huu-Thuan} \snm{Huynh}}
			\author[
			addressref={aff1},
			email={nakashim@is.naist.jp}
			]{\inits{YN}\fnm{Yasuhiko} \snm{Nakashima}}
			
			
			\address[id=aff1]{
				\orgname{Nara Institute of Science and Technology}, 
				\street{Takayama 8916-5},                     %
				\postcode{630-0101}                                
				\city{Ikoma, Nara},                              
				\cny{Japan}                                    
			}
			\address[id=aff2]{%
				\orgname{Vietnam National University - Ho Chi Minh City, University of Science},
				\street{227 Nguyen Van Cu Str., Wd.4, Dist.5},
				\postcode{}
				\city{Ho Chi Minh City},
				\cny{Vietnam}
			}
			
			
			\begin{artnotes}
			\end{artnotes}
			
			
			
			\begin{abstractbox}
				
				\begin{abstract} 
					Visible Light Communication (VLC)-based systems are now applied widely in indoor positioning systems (IPS) in which VLC-based LED-beacons are assigned with fixed position identifications. In IPS, massive installation efforts of micro-controllers or programmable oscillators for VLC-LED beacons will be cut down if VLC-specialized hardwares are applied. Unfortunately, hardware implementations of VLC transmitters and receivers are not investigated until now; and encoding/decoding of run-length limited (RLL) codes and forward error correction (FEC) codes are processed purely on firmwares of embedded processors. However, recent works on soft-decoding of RLL have shown that they are heavy and time-consuming computations. This could be an obstacle, especially in low-end micro-controllers (MCUs) or system on chips (SoCs). In this paper, we introduce a couple of hardware implementations of compact VLC transmitter and receiver for the first time. Compared with related works, our VLC transmitter is non-RLL one, that means flicker mitigation can be guaranteed even without RLL codes. In particular, we have utilized a centralized bit probability distribution of a prescrambler and a Polar encoder to create a non-RLL flicker mitigation solution. Moreover, at the receiver, a 3-bit soft-decision filter is proposed to analyze signals received from real VLC channel to extract log-likelihood ratio (LLR) values and feed them to the FEC decoder. Therefore, soft-decoding of Polar decoder can be implemented to improve the bit-error-rate (BER) performance of the VLC system. Finally, we introduce a novel very large scale integration (VLSI) architecture for the compact VLC transmitter and receiver; and synthesis our design under FPGA/ASIC synthesis tools. Due to the non-RLL basic, our system has an evidently good code-rate and a reduced-complexity compared with other RLL-based receiver works. Also, we present FPGA and ASIC synthesis results of the proposed architecture with evaluations of power consumption, area, energy-per-bits and so on. 
				\end{abstract}
				
				
				
				
			\end{abstractbox}
		\end{fmbox}
		
	\end{frontmatter}



\section{Introduction}
\label{intro}

\subsection{VLC-beacon-based indoor positioning systems (IPS)}
\label{intro-beacon}

VLC simultaneously provides both illumination and communication services. Specifically, VLC systems currently utilize visible light for communication that occupy the 380nm-750nm spectrum \cite{Mostafa,Latif}. Some modulation schemes have been introduced for VLC systems, e.g. Variable Pulse Position Modulation (VPPM), On-off Keying (OOK), or Orthogonal Frequency Division Multiplexing (OFDM) and so on \cite{Latif,Junhai}. The VLC transmitter modulates the digital information to light signals through  a transmit (TX) front-end and a light-emitting diode (LED). 
Generally, indoor localization applications which show users' locations in indoor buildings are getting more attentions from researchers and industry in recent years \cite{Junhai}. Several statistics show that human spend almost 80\% time of a day indoor where global positioning systems (GPS) could not work \cite{Lifang2}. Accordingly, indoor localization is the key to open a wide range of location-based service (LBS) applications. Indeed, mobile indoor positioning in retail is estimated up to \$5 billion in 2018 \cite{Junhai}. Current approaches in indoor positioning which are often based on Wi-Fi, Ultra-wideband (UWB), Radio-Frequency Identification (RFID), or other RF wireless techniques \cite{Junhai}. These approaches often meet problems related to high cost of installation and management; or can not be used in Radio Frequency (RF) banned areas such as hospitals, planes or gas stations \cite{Junhai}. VLC-based indoor positioning solutions have promising characteristics such as low cost, high security, high spatial reuse, low co-channel interference, high-precision and so on \cite{Lifang2,Junhai}. VLC-based solutions, therefore are considered widely as suitable candidates for indoor positioning. In VLC-beacon-based indoor localization systems, unique ID information are transmitted from VLC-LED bulbs for purposes such as identifying objects and locations \cite{Yoshizawa}. Furthermore, beacon-based frames have been introduced in some publications with the sizes of 158-bit \cite{Yoshizawa}, 56-bit \cite{Qing} or 34 symbols (0.96ms) \cite{Yamazato}. We found that the 158-bit beacon-based frame which is defined by Standard of Japan Electronics and Information Technology Industries Association (JEITA) \cite{Latif,Yamazato} should be considered because this work is confirmed by an association. Particularly, the structure of the JEITA's beacon-based frame includes three parts: start of frame (SOF), payload, and the end of frame (EOF). The SOF includes 6-bit preamble indicating the beginning of the frame, and another 8-bit defines the frame type. The payload includes 128-bit ID data. Finally, 16-bit cyclic redundancy is reserved for error correction \cite{Yoshizawa}. 

There is one fact that beacon broadcasting of VLC-based indoor localization systems does not require a high-speed link. Therefore, throughout this paper, we consider the OOK modulation because of its simplicity and easy implementation. Also, we favor in setting a low frequency for the proposed system to evaluate its performance.

\begin{figure*}[h!]
	\centering
	\includegraphics[width=3.5in]{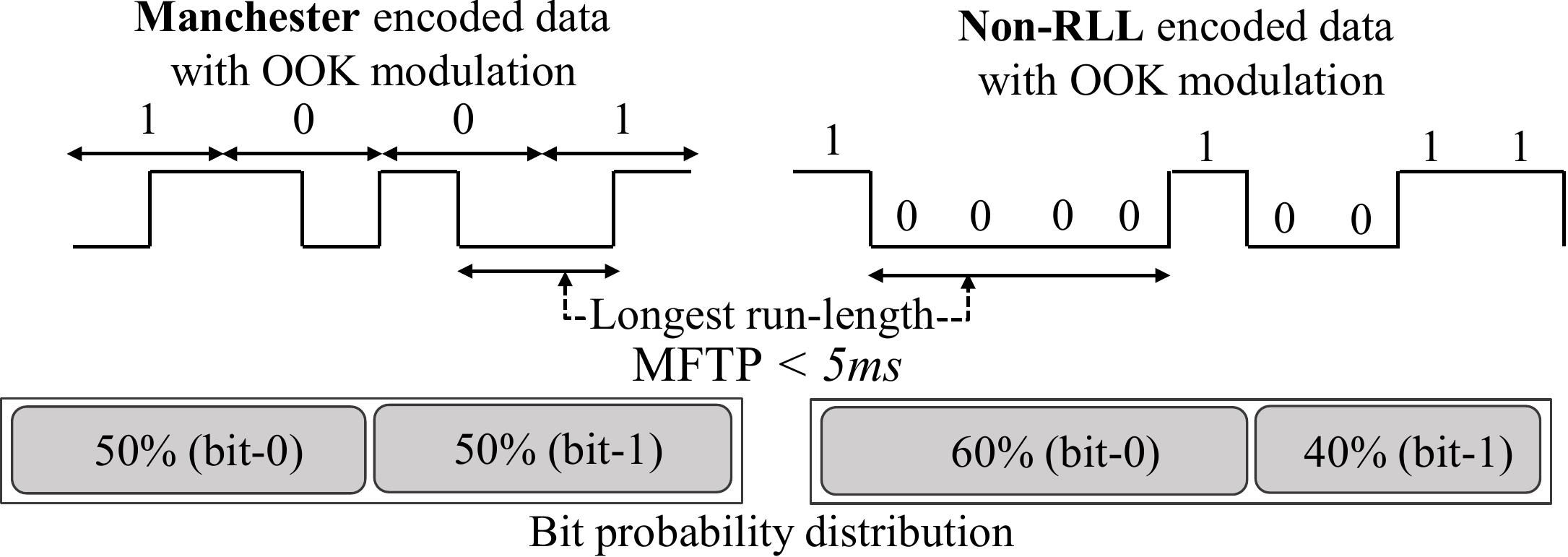}
	\caption{\csentence{}
		Run-length, bit probability distribution and flicker mitigation}
	\label{fig1}
\end{figure*}

\subsection{Flicker mitigation problem}
\label{flicker}
The brightness and stability of the light are strongly affected by the distribution of the 1\char`\'s and 0\char`\'s in the data frames. RLL coding is indispensable to avoid LED's flicker and guarantee the direct current (DC) balance in visible light communication systems. Therefore, many DC-balance solutions are introduced to maintain approximately equal numbers of zero and one bits in the data frames. As a result, flicker mitigation which based on DC-balance techniques is considered as one of essential concerns in any VLC systems. Moreover, when the light source is modulated for data communication, run-length of the data codewords should be carefully controlled to mitigate the potential flickers. To avoid flicker, the changes in brightness must be faster than the maximum flickering time period (MFTP), which is defined by the maximum time period that light intensity can change without being perceived by human eyes \cite{Fang}. In normal cases, a MFTP which is faster than 5 ms is considered safe for a non-flicker guarantee. Fig.\ref{fig1} shows an illustrative example to introduce how run-length and bit probability distribution affect to the flicker of VLC systems in case of light is modulated by OOK method. When data is modulated by Manchester coding, the maximum run-length is limited to 2 while the ratio of bit-0 and bit-1 are always equal in all cases. On the contrary, bit-distribution and run-length of non-RLL cases are arbitrary. Therefore, non-RLL approaches potentially cause flickers which could be recognized at the LED bulbs. As a result, whenever the non-RLL scheme is considered for VLC systems, the run-length and centralized bit probability distribution should be carefully investigated. Also, the lowest transmit frequency that can guarantees flicker mitigation should be considered in such non-RLL OOK VLC systems.

\subsection{Why the hardware implementation of VLC transmitter and receiver is important?}
\label{whyHW}

\begin{figure*}[h!]
	\centering
	\includegraphics[width=3.5in]{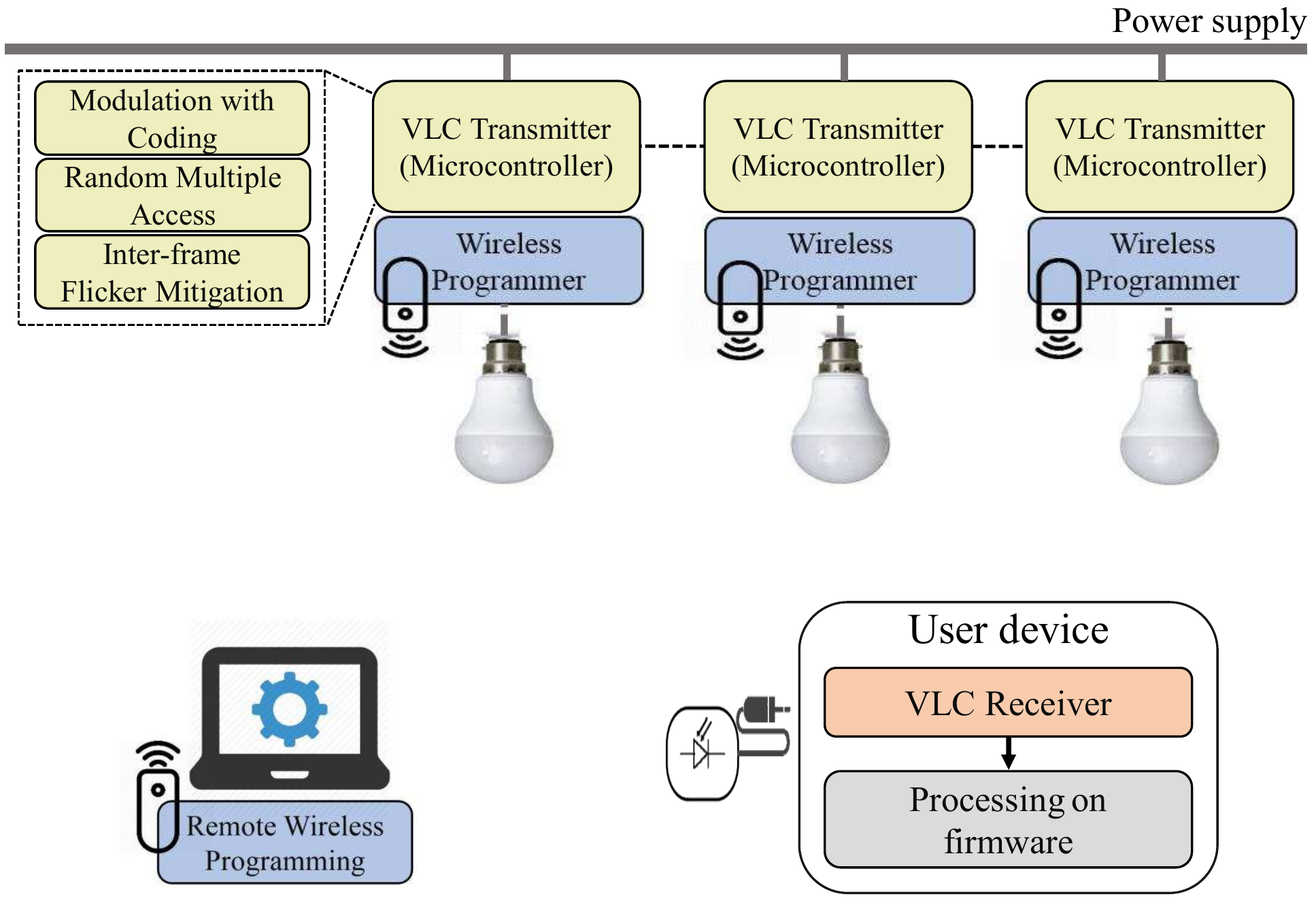}
	\caption{\csentence{}
		An example of a VLC-based indoor positioning system}
	\label{fig2}
\end{figure*}

Fig.\ref{fig2} shows an example of a typical VLC-based indoor localization system in which VLC transmitter's function blocks are mainly processed by a firmware program on a trivial micro-controller; while the VLC receiver and positioning algorithm are executed on a firmware program of a user's portable device \cite{Qing}. Furthermore, an optional part of the VLC transmit (TX) package is the wireless programmer which helps configure the firmware on the low-end micro-controller remotely. It can be found that when VLC-based indoor positioning system is applied inside a large building, in which hundreds or thousands of VLC-LED bulbs required. In this case, the implementation cost increase linearly because each micro-controller is dedicated for only one VLC-LED anchor \cite{Qing}, or several LEDs \cite{Kuo}. Moreover, each VLC-beacon package takes more space to integrate the programing circuits. On the other hand, if we assume that only one micro-controller is employed to control many VLC-LED beacons and long wires are used for routing to LED bulbs through VLC TX front-ends. As a result, encoding of FEC and RLL codes is processed sequentially on MCU's firmware before encoded data is feed to numerous of VLC TX beacons. Although, FEC encoding or RLL encoding are not time-consuming tasks. However, to due with a large number of VLC transmitters, this scenario sometimes limits the smooth operation of the VLC-based beacon system in which its time constraints and flicker mitigation must be guaranteed. Moreover, due to the limited capability of the low-end micro-controller, only a few VLC-based beacons are well managed by one micro-controller. Hence, this restricts the scalability of the VLC-based indoor localization systems.     

On the contrary, VLC receiver's function blocks which include decoding of RLL and FEC codes are purely processed on user's portable device. However, some soft-decoding algorithms of RLL and FEC have been proposed recently in VLC systems\cite{Wang2,Wang3,Wang4,Le}. These solutions help improve the performance of the VLC receiver; however, they are potentially time-consuming tasks with many complex mathematic computations. Besides, fastidious users always expect real-time responses or for their indoor positioning application. Therefore, VLC soft-decoding receivers or localization algorithms need to be optimized or     

\begin{figure*}[h!]
	\centering
	\includegraphics[width=4in]{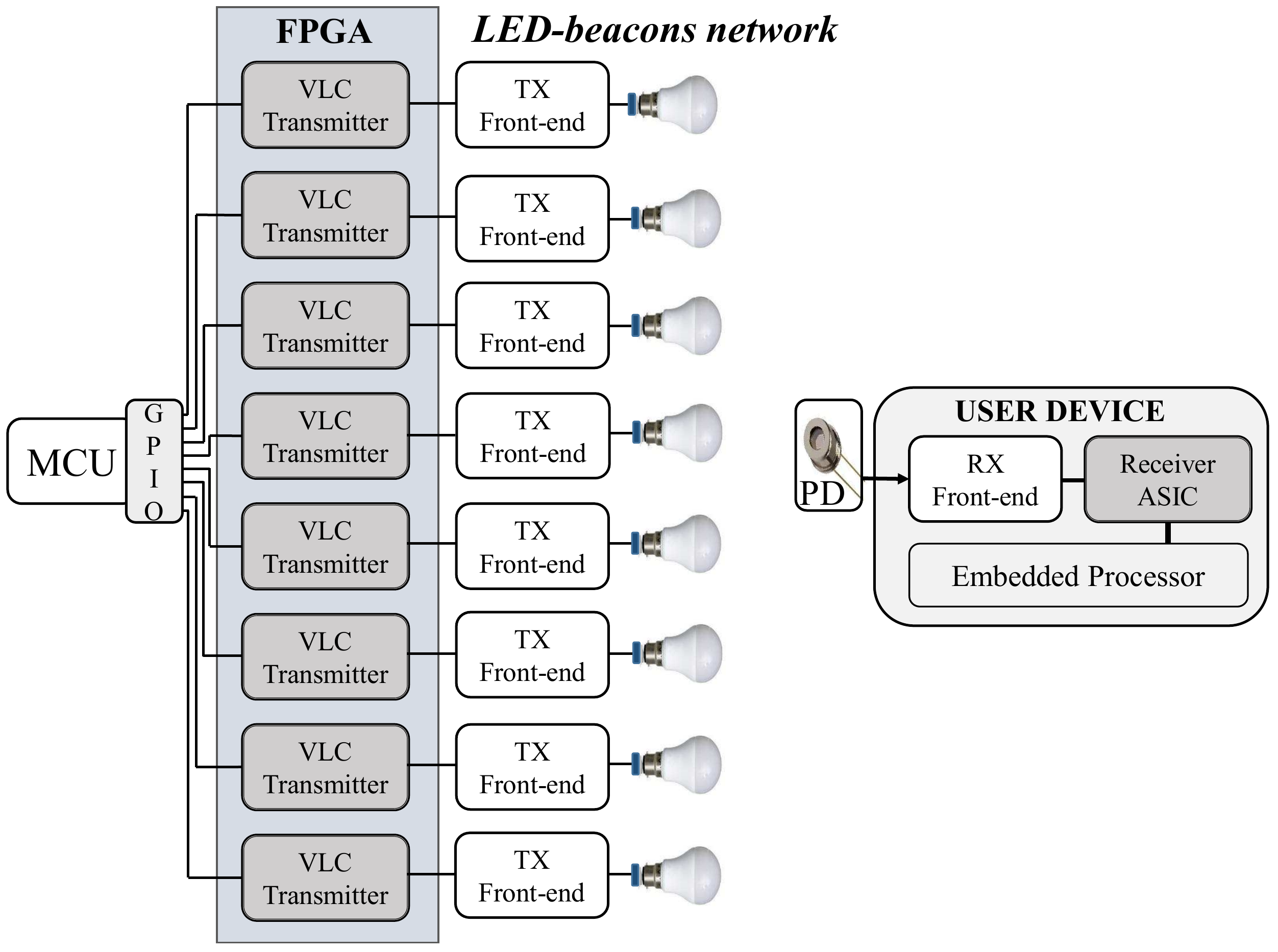}
	\caption{\csentence{}
	The proposed VLC-LEDs beacons network based on our hardware works}
	\label{fig3}
\end{figure*}

Regarding with reasons mentioned above, we have proposed two dedicated hardware implementations of VLC transmitter and receiver. The overview of our proposal is briefly presented in Fig.\ref{fig3}. In particular, we have utilized the parallel processing capability of the FPGA to implement VLC transmitters inside an FPGA which connects to many TX front-ends in the LED-beacon network. Specifically, one VLC transmitter hardware executes tasks, for instance, modulation with coding, random multiple access or inter-frame flicker mitigation \cite{Qing}. Accordingly, ID information of each VLC-LED bulb is processed directly at each VLC transmitter right after the center MCU pushes coarse bits to GPIO ports. Therefore, only one micro-controller is required to monitor all the IDs issued for all LED bulbs. At the user device, the dedicated VLC receiver ASIC is expected to enhance the processing time of soft-decoding of FEC or RLL codes, which contains heavy mathematic computations e.g. multiplication, exponential, logarithm functions and so on \cite{Le,Wang3}. Hence, VLC-based indoor localization systems can be operated smoothly without recognizable delays. In this paper, we introduce a couple of hardware implementations with VLSI architectures of the proposed compact VLC transmitter/receiver; in which essential problems related to flicker mitigation and soft-decoding of RLL and FEC are clearly discussed.  

\section{Related works}
\label{relatedwork}

Table \ref{table_1} summarizes proposals related to FEC and flicker mitigation for VLC. The conventional solution is defined in the IEEE 802.15.7 standard, which employs Reed-Solomon (RS) codes, Convolutional Codes (CC) and RLL codes with hard-decoding of RLL codes (hard-RLL) \cite{Sridhar}. However, hard-RLL methods of inner RLL codes limit to hard-decoding of outer FEC codes \cite{Sridhar,Wang1,Sunghwan}; consequently, the error-correction performance of the entire VLC system is restricted. Recently, soft-decoding RLL (soft-RLL) solutions have been proposed in \cite{Wang2,Wang3,Wang4,Le}. These techniques permit soft-decoding FEC algorithms to be applied to improve the bit-error-rate (BER) performance of the VLC system, but they also require heavy computational efforts, with many additions and multiplications. 

Zunaira \emph{et al.} have proposed replacing the classic RLL codes with a recursive Unity-Rate Code (URC) or an Unary-Code as the inner code, and a 17-subcode IRregular Convolutional Code (IRCC) is selected for the outer code \cite{Babar,babar2018unary}. Although these methods can achieve different dimming levels with good BER performances; however, the system latency is increased with the iterative-decoding schemes. In addition, the reported codeword length is rarely long, which ranges from 1000 to 5000 bits, reduces the compatibility of this proposal to VLC-based beacon systems \cite{Yoshizawa,Qing} in which beacon-based frame sizes are always small. As an alternative approach, Kim \emph{et al.} have proposed a coding scheme based on modified Reed-Muller (RM) codes \cite{Sunghwan}. Although this method can guarantee DC balance at exactly 50\%, it has the inherent drawbacks of a deducted code rate and an inferior error-correction performance compared with turbo codes, low-density parity-check (LDPC) codes or polar codes. In addition, Lee and Kwon have proposed the use of puncturing and pseudo-noise sequence scrambling with compensation symbols (CS) \cite{Lee}. This proposal can achieve very good BER performance; however, puncturing with CSs will lead to redundant bits in the messages, thereby reducing the transmission efficiency. Another coding scheme based on the fountain code, which has greatly improved the transmission efficiency, is mentioned in \cite{Lifang}. However, this scheme requires feedback information and thus is not suitable for broadcasting scenarios in VLC-based beacon systems. Xuanxuan Lu \emph{et al.} have reported a new class of enhanced Miller codes, termed eMiller codes which is a class of RLL codes known for high-bandwidth efficiency \cite{Xuanxuan}. Besides, she also proposed an improved version of Viterbi algorithm, termed \emph{mnVA} to further enhance the performance of her proposed eMiller code. It can be seen from her simulation results that eMiller helps improve the performance of the whole VLC system; and this code seems to be a promising candidate for VLC applications. However, we have found two main drawbacks of this approach are the unoptimized code-rate = 1/2 of the eMiller code (Table \ref{table_2}), and an increasing in computational complexity.   
   
\begin{table}[h!]
	\caption{Overview of FEC algorithms and flicker mitigation solutions for VLC}
	\label{table_1}
	\begin{tabular}{cc}
		\hline
		FEC solution & Flicker mitigation \\
		\hline
		RS, CC \cite{Sridhar} & Hard-RLL \\
		Multi-RS hard-decoding \cite{Wang1} & Hard-RLL\\
		LDPC \cite{Sunghwan1} & Hard-RLL \\
		RS soft-decoding \cite{Wang2,Wang3} &Soft-RLL \\
		Polar code \cite{Wang4,Le} & Soft-RLL \\
		Irregular CC \cite{Babar} & Unity-Rate Code\\
		Irregular CC \cite{babar2018unary} & Unary-Rate Code\\
		Reed-Muller \cite{Sunghwan}& Modified original code\\
		Turbo code \cite{Lee} & Puncture + Scrambling\\
		Fountain code \cite{Lifang} & Scrambling\\
		Convolutional code, Viterbi \cite{Xuanxuan} & Enhanced Miller code\\
		Polar code ($N$=2048) \cite{Fang}& Flicker-free\\
		Proposed method ($K$=158, $N$=256 )& Flicker-free\\
		(JEITA's beacon frame size) & \\
		\hline
	\end{tabular}
\end{table}

Advantages of Polar code are exploited deeply together with soft-decoding of RLL codes have been introduced at \cite{Wang4,Le}. According to these publications, Manchester and 4B6B codes are used as RLL solutions for the VLC system. As a result, their BER performances have been improved remarkably with a flexibility of Polar code's code-rate. However, we found that the code-rate = 1/2 of Manchester code, or code-rate = 0.67 of 4B6B (summarized at the Table \ref{table_2}) are also not the best optimization solution for channel efficiency enhancement, if compared with non-RLL approaches. Fang\emph{et al.} have recently proposed a non-RLL polar-code-based solution for dimmable VLC systems \cite{Fang}. This approach has shown promising results in weight distribution and run-length distribution. Moreover, this solution also shows an improved transmission efficiency while achieving a high coding gain compared with RS and LDPC codes. We have found that this solution can overcome most of the drawbacks of the related works mentioned until now. Specifically, it offers the non-iterative decoding with a low-complexity. Also, it has a flexible code-rate, and a high BER performance without requiring any feedback information. However, we found that the biggest obstacle of this proposal is the equal probabilities of short runs of 1\char`\'s and 0\char`\'s can only be achieved with a long codeword length; as chosen to be \textit{N}=2048. Indeed, long data frames rarely be applied in low-throughput VLC systems, for instances, VLC-based beacon ones \cite{Yoshizawa,Qing}. It can be found that the non-RLL solution based only on a polar encoder \cite{Fang} might not be applicable in such VLC-based beacon systems because DC balance is not guaranteed for short data frames. 

In the later parts of this paper, we point out the unsolved problems of non-RLL flicker mitigation in VLC-based beacon systems. Additionally, as mentioned in Section \ref{whyHW}, we introduce a couple of non-RLL beacon-based VLC transmitter and receiver and their VLSI architectures for the first time. In summary, our contributions include:
\begin{enumerate}
	\item First discussion on the importance of FPGA and ASIC implementations of VLC transmitters and receivers in VLC-based beacon systems (Section \ref{whyHW})
	\item A non-RLL flicker mitigation method based on a prescrambled Polar encoder (Section \ref{flickermitigation}).
	\item Two proposed hardware architectures of beacon-based VLC transmitter and receiver (Section \ref{hwarchitecture}).
	\item A 3-bit soft-decision filter which can support soft-decoding of FEC decoders in real prototypes of VLC receivers (Section \ref{filter}).
\end{enumerate}
    
\section{Flicker mitigation based on a non-RLL prescrambled Polar encoder}
\label{flickermitigation}

\begin{figure*}[h!]
	\centering
	\includegraphics[width=4in]{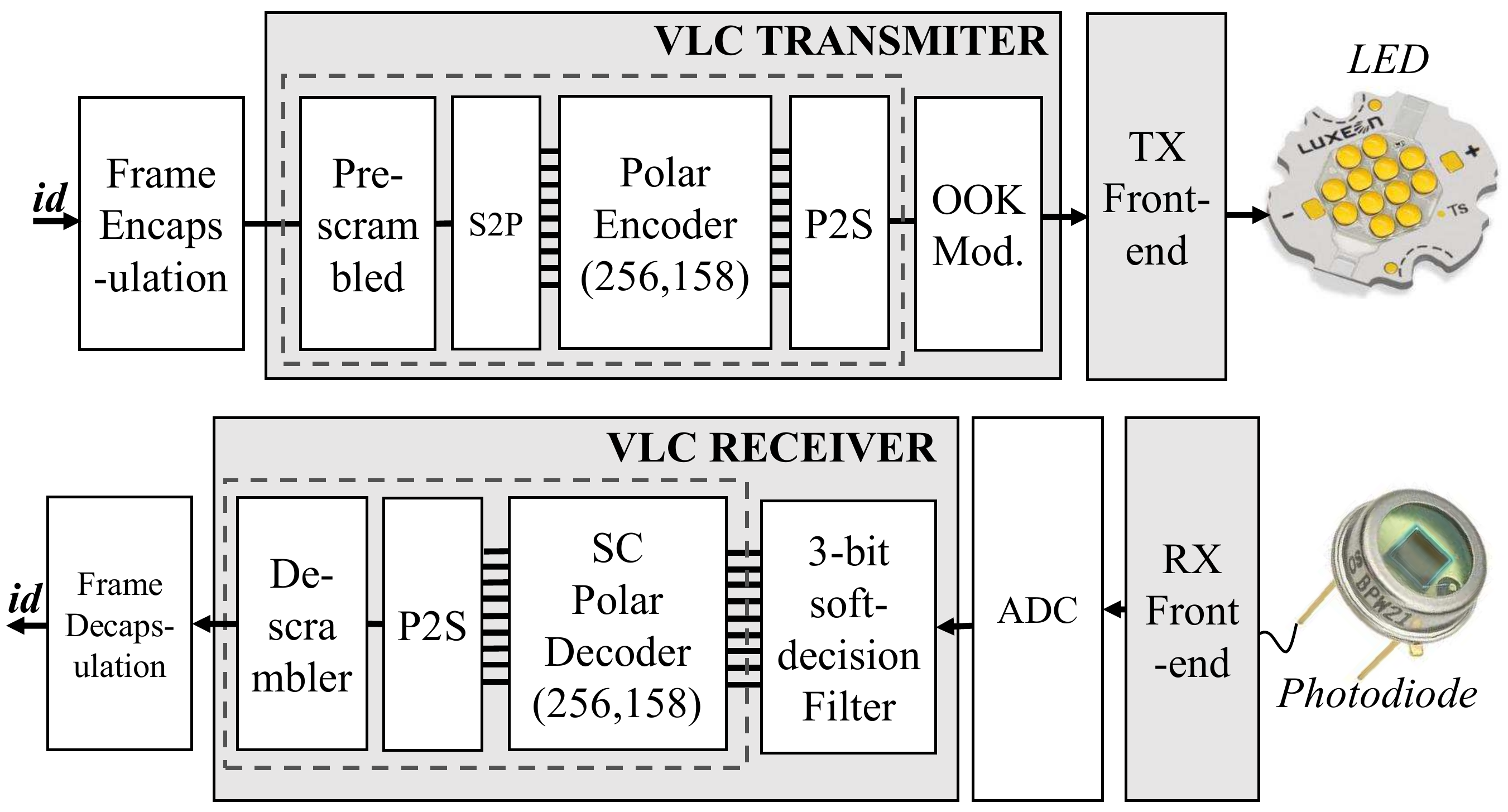}
	\caption{\csentence{}
		Block diagram of the proposed VLC transmitter/receiver hardware architecture}
	\label{fig4}
\end{figure*}

It follows from the Section \ref{relatedwork}, due to the small size of beacon-based data frames, a non-RLL DC-balance solution which dedicated for the VLC-based beacon systems seems still to be an unsolved problem. In this section, we introduce a non-RLL flicker mitigation solution which is designed for VLC-based beacon systems. Particularly, our flicker mitigation solution is the combination of a simple pre-scrambler placed at the outer code, with a (256;158) polar encoder placed at a inner code's position. Fig. \ref{fig4} briefly introduces our proposal in style of block diagram.  

Table \ref{table_2} summarizes a code-rate comparison of RLL and non-RLL solutions. It can be noticed that non-RLL solutions keep the system rate unchanged while removing the heaviness of RLL encode/decode blocks. Furthermore, FEC decoders also inherit from the removing RLL codes because soft-decoding of them can be implemented without difficulties in achieving LLR values. However, DC-balance and run-length should be controlled strictly in such non-RLL VLC systems.    

\begin{table}[h!]
	\caption{Code-rate comparison of non-RLL and RLL solutions}
	\label{table_2}
	\begin{tabular}{cc}
		\hline
		Code & Code-rate \\
		\hline
		Manchester & 1/2 	    	\\
		FM0/FM1    & 1/2		    \\
		Conventional Miller & 1/2   \\
		eMiller [15]      & 1/2     \\
		4B6B      & 0.67		   \\
		8B10B      & 0.8 			\\
		non-RLL (our work)      & 1 (No changed) \\
		\hline
	\end{tabular}
\end{table}

In a digital transmission system, a data scrambler plays an important role because it causes energy to be spread more uniformly. At the transmitter, a pseudorandom cipher sequence is modulo-2 added to the data sequence to produce a scrambled data sequence. 

Describe the generating polynomial P(x) as:
\begin{equation}\label{eq:polynomial}
P(x)=\displaystyle\sum_{q=0}^{N} c_q.x^q
\end{equation}
where $c_0$ = 1 and is equal 0 or 1 for other indexes.

We have found that the output bit probability distributions of pre-scramblers in different generating polynomials seem to differ slightly. Therefore, we propose a simple generating polynomial presented in (\ref{eq:scrambler}) to reduce the number of shift registers required for a pre-scrambler.  
\begin{equation}\label{eq:scrambler}
P(x)=x^4 + x^3 + 1
\end{equation}

Meanwhile, polar codes can be classified into two types: non-systematic and systematic codes. Typically, a polar code is specified by a triple consisting of three parameters: \textit{(N, K, I)}, where \textit{N} is the codeword length, \textit{K} is the message length, and \textit{I} is the set of information bit indices. Let \textit{d} be a vector of \textit{N} bits, including information bits. The generator matrix is defined as \(G=(F^{\otimes n})_I\). Then, given a scrambled message \textit{u} of \textit{K} bits in length, a codeword \textit{x} is generated as given in (\ref{eq:polarencode}). 
\begin{equation}\label{eq:polarencode}
x = u.G = d.F^{\otimes n}
\end{equation}

A Polar encoder is formed of many layers of XOR gates, with a complexity of \(\frac{N}{2}log_2N\) XORs. There is one fact that systematic polar codes were introduced to achieve better error-correction performances compared with non-systematic codes \cite{Harish}. However, due to the information bits transparently appear as a part of the codeword, we have found that the output bit-probability distribution of a systematic Polar encoder (SPE) is not well centralized. On the other hand, the output bit probability distribution of an non-systematic Polar encoder (NSPE) naturally becomes centralized approximately 50\% 1\char`\'s and 50\% 0\char`\'s when the codeword length is long enough \cite{Fang}. 

In summary, we have selected the Polar code as the main FEC scheme for our VLC-based transmitter/receiver due to several reasons:
\begin{enumerate}
	\item The encoder's output bit probability distribution is naturally centralized when long codewords are applied in the system.
	\item Unusual code rates are supported. Specifically, a (256;158) polar code, which has a code rate of 0.617, is suitable for a beacon-based frame size of \textit{K}=158.
	\item High error-correction performance can be achieved with a low hardware complexity \cite{Phuc}.
	\item The inherently short run lengths of a polar encoder can be useful in mitigating the lighting flicker \cite{Fang}.
\end{enumerate}
A pre-scrambler can help to ensure the fast convergence of the output probability distribution of an inner (256;158) Polar encoder. As a result, DC balance in a VLC-based beacon system can be guaranteed by the proposed transmitter depicted in Fig.\ref{fig5}. 

\section{Hardware architecture of the proposed VLC transmitter and receiver}
\label{hwarchitecture}

\subsection{Hardware architecture of the VLC transmitter}
\label{transmitterarch}

\begin{figure*}[h!]
	\centering
	\includegraphics[width=3.5in]{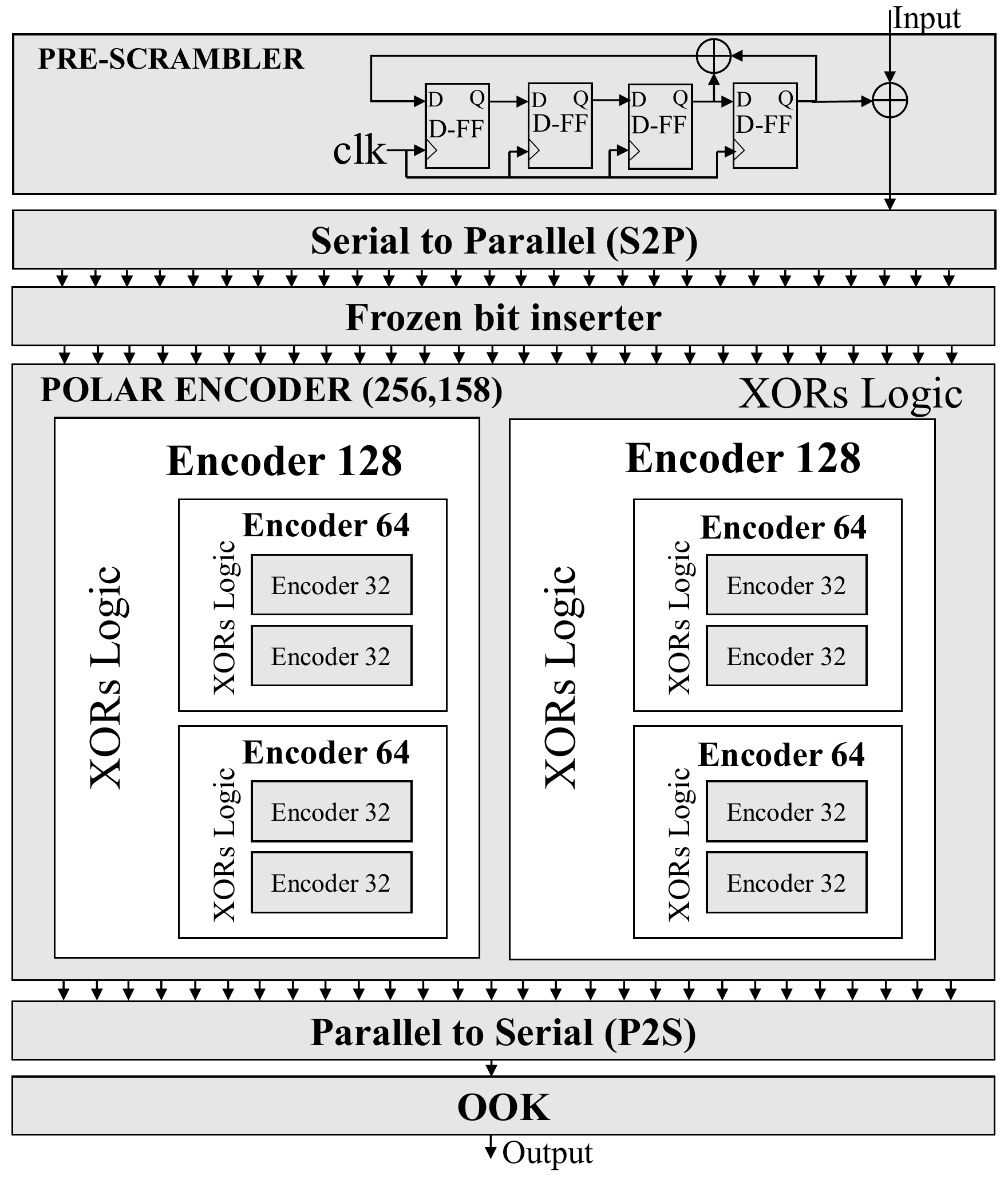}
	\caption{\csentence{}
		The hardware architecture of our non-RLL VLC transmitter}
	\label{fig5}
\end{figure*}

Block diagram of the proposed VLC transmitter is shown in Fig.\ref{fig5}. As mentioned in Section \ref{relatedwork}, it seems that Polar code is an optimal candidate for a FEC solution in VLC receiver \cite{Fang,Le}. In Section \ref{flickermitigation}, we have also introduced a pre-scrambled Polar encoder as a non-RLL flicker mitigation in case of beacon-sized codewords which defined by JEITA are applied in the VLC-based beacon systems. In fact, the IEEE 802.15.7 standard has stated that Reed-Solomon (RS) and convolutional codes are preferred over low density parity check (LDPC) codes in order to support short data frames, hard-decoding with low complexity \cite{Sridhar}. We found that flexible code-rates of the Polar code can support any sizes of data frames \cite{Harish}. Also, its soft-decision decoding can improve the reliability of the VLC systems compared with RS and convolutional codes. Moreover, the inherent low-complexity characteristic of Polar code's encoding and Successive-Cancellation (SC) decoding is suitable for being applied in VLC receivers.

In the proposed VLC transmitter described in Fig.\ref{fig4} Fig.\ref{fig5}. Firstly, 128-bit ID information data is wrapped by a frame encapsulation procedure to form a 158-bit beacon-based frame \cite{Yoshizawa}. Next, the 158-bit frame is scrambled by a pre-scrambler. Due to a simple generating polynomial (\ref{eq:scrambler}) is applied, only four registers and one XOR gate are required to create a pre-scrambler. The frozen bit inserter plays a role of inserting $N-K$ frozen bit indices into different positions of a 256-bit frame. In particular, in case JEITA's 158-bit beacon-based frame is applied, 98 frozen-bits are inserted at positions defined at the Polar code construction stage. After frozen bits are inserted, the pre-scrambled 256-bit frame is encoded by a Polar encoder (256;158) to create a bit stream in which the DC-balance can be guaranteed even without any RLL codes. Regarding with the Polar encoder, we have implemented a recursive combinational architecture for the Polar encoder, in which $2^{N}$-code-length encoders are created by $N/2$ XOR gates and two $2^{N-1}$-code-length encoders Fig.\ref{fig5}. Due to the block encoding characteristic of Polar encoder, the Serial-to-Parallel (S2P) block is implemented to prepare the pre-scrambled serial bit-stream to a 256-bit register. This register is the input register of Polar encoder. Also, Parallel-to-Serial (P2S) block converts parallel Polar encoded bits to serial bit stream before being modulated by the OOK block. Finally, the VLC TX front-end converts the OOK-modulated signals to light signals and broadcast them to the air. Specifically, we have also assembled a VLC TX front-end that successfully transmit information through a normal 5V LED with a transmit frequency up to 2.5 Mhz. 

\begin{figure*}[h!]
	\centering
	\includegraphics[width=3.5in]{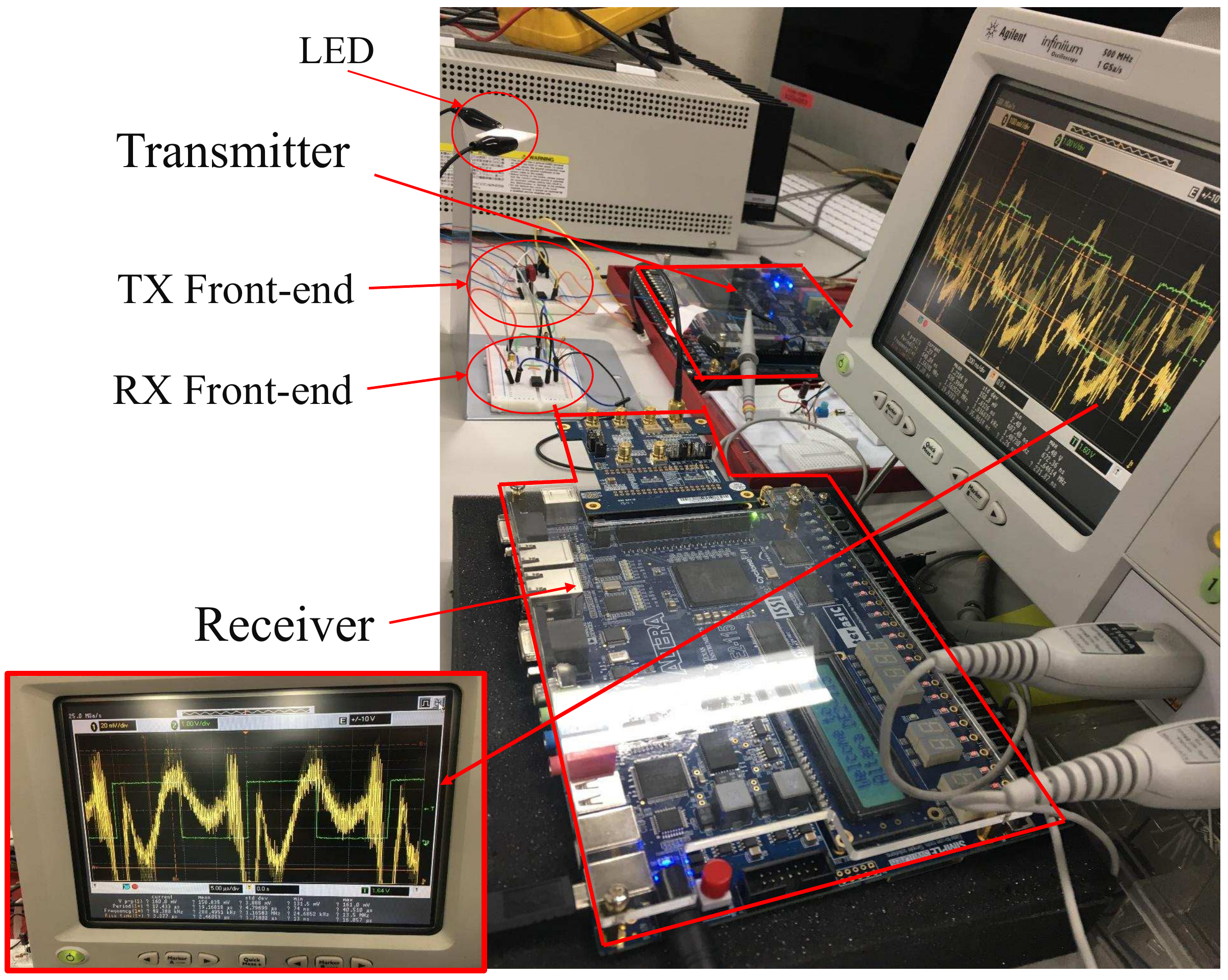}
	\caption{\csentence{}
		Distorted received signals due to the bad channel settings}
	\label{fig6}
\end{figure*}

\subsection{Hardware architecture of the VLC receiver}
\label{receiver}

\subsubsection{3-bit Soft-Decision Filter}
\label{filter}

\begin{figure*}[h!]
	\centering
	\includegraphics[width=3.5in]{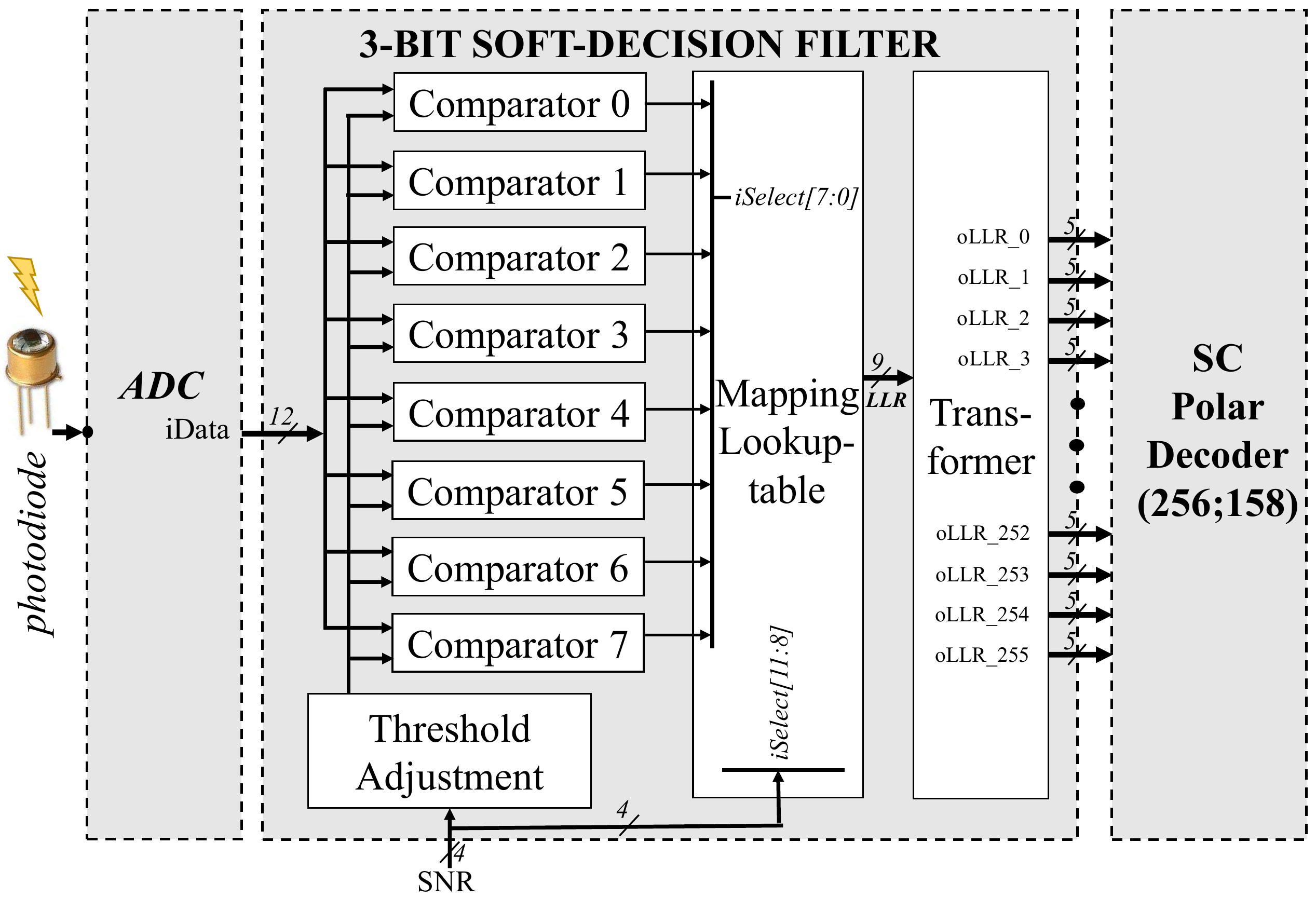}
	\caption{\csentence{}
		Hardware architecture of the 3-bit soft-decision filter}
	\label{fig7}
\end{figure*}

Fig.\ref{fig6} shows our FPGA-based VLC demonstration system in which distorted signals are received at the VLC RX front-end, then it is displayed on the oscilloscope. Specifically, we have found that distortions appear in two experimental scenarios. Firstly, when the transmit frequency is higher than the maximum frequency that RX front-end can receive. Secondly, when the distance between TX LED and RX front-end increases in space. Also, distortions of the received signals also appear with shrunken peak-to-peak voltages (\textit{Vpp}). Distorted received signals are usually the cases cause reliability of the VLC system deducted because hard-decoding of RLL and FEC are often the default selections in most VLC receivers \cite{Sridhar}. In this paper, we introduce a 3-bit soft-decision filter which is implemented at VLC receiver to support soft-decoding of RLL and FEC decoders in real VLC receiver prototypes.   

Specifically, in the case of VLC AWGN channel, a sequence of the \textit{LLR} values which is necessary for soft-decoding of FEC decoder, are expressed by Eq.\ref{eq4}. 
\begin{equation}
LLR(y_i) = \ln{\frac{P(x_i=0|y_i)}{P(x_i=1|y_i)}}
\label{eq4}
\end{equation}
where $y_i$ is the received sample and the conditional probability is generally calculated as Eq.\ref{eq5}.
\begin{equation}            
P(x_i|y_i = \Delta) = \frac{1}{\sqrt{2\pi\sigma_{\Delta}^{2}}}e^{-\frac{(y_i-\mu_{\Delta})^2}{2\sigma_{\Delta}^{2}}}
\label{eq5}
\end{equation}  

where $\mu_{\Delta}$ and $\sigma_{\Delta}$ are the mean value and standard deviation for $\Delta = 0, 1$. However, when making real prototype of soft-decoding VLC receiver, we found that it is unfeasible in estimating the LLRs using such Eq.(4) and Eq.(5) due to $\mu_{\Delta}$ and $\sigma_{\Delta}$ can not be estimated in real optical wireless channels. Therefore, in this paper, we propose applying a soft-decision filter which is first introduced in optical communication systems for our VLC receiver prototype \cite{Tagami}. 

\begin{figure*}[h!]
	\centering
	\includegraphics[width=4in]{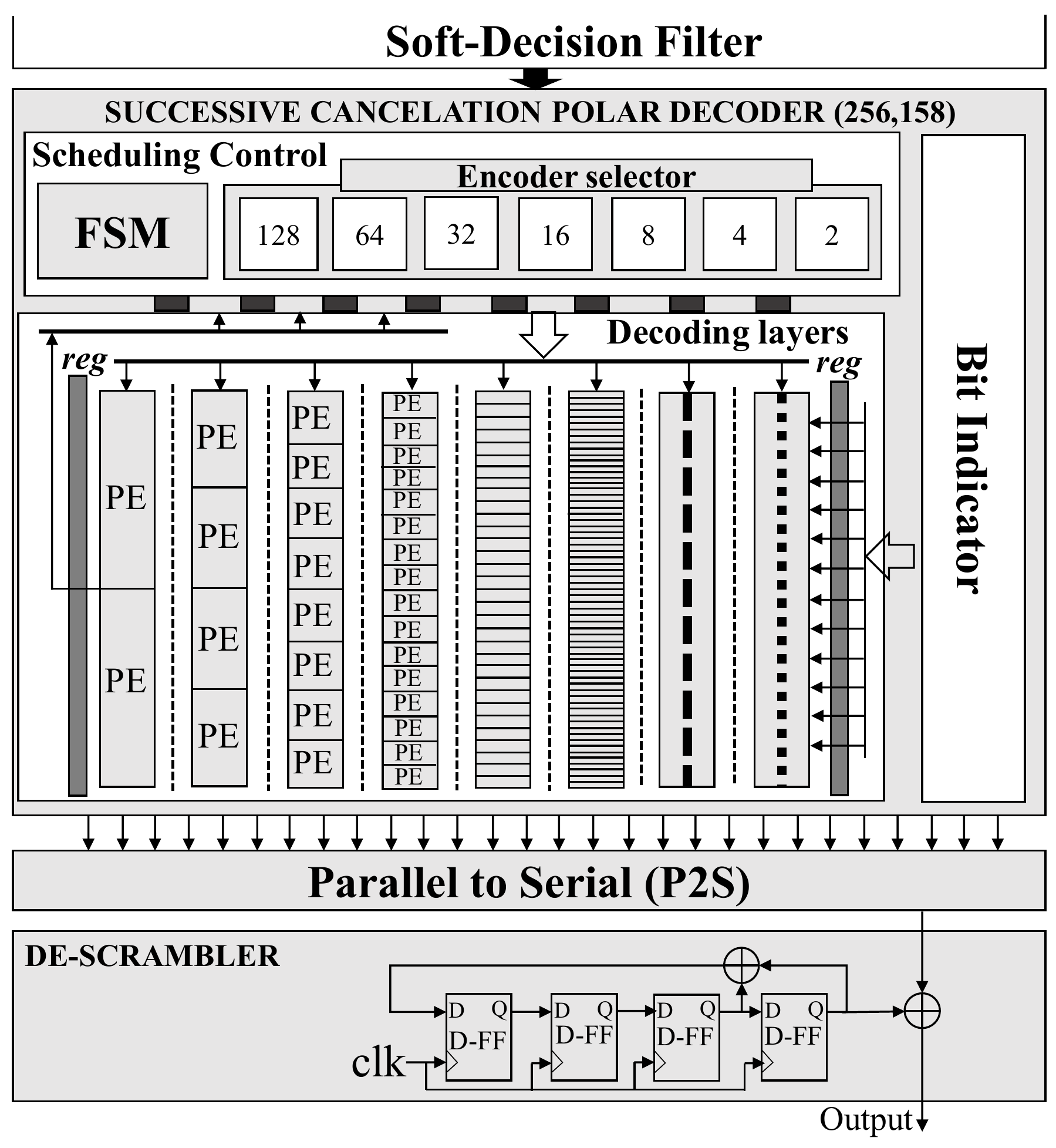}
	\caption{\csentence{}
		The hardware architecture of our non-RLL VLC receiver}
	\label{fig8}
\end{figure*}

Fig.\ref{fig7} shows the proposed hardware architecture of 3-bit decision filter that we have implemented. Firstly, an analog-to-digital converter (ADC) converts analog signals received from the RX front-end to digital signals. The 3-bit soft-decision filter analyses these digital signals and calculate LLR values to feed to soft-decoding Polar decoder. The soft-decision filter includes $2^{N-1}$ decision thresholds to compare with the incoming received signal, where \textit{N} is the number of quantization bits. Previous research on soft-decision filter in optical communication systems has shown that 3-bit soft decision was the optimum solution \cite{Tagami}. In the case of \textit{N=3} for 3-bit soft decision, we established seven threshold voltages from $V_{t+3}$ to $V_{t-3}$ which are calculated from equations given in Eq.\ref{thresholdVoltage}. We have defined a mapping table with output LLR values are carefully chosen from training simulation results on MATLAB. Table \ref{table_3} shows ranges of comparison and their output LLR values. 
The sequence of 9-bit LLR results of mapping lookup-table are buffered and quantized by a block named \textit{Transformer}, before passing them to the Polar decoder.

\begin{equation}
\label{thresholdVoltage}
\begin{split}
& V_{t} = \frac{V_{peak+} + V_{peak-}}{2} \\
& V_{t-1} = V_{t} - \frac{V_{peak+} - V_{t}}{4} \\
& V_{t-2} = V_{t} - 2\frac{V_{peak+} - V_{t}}{4} \\
& V_{t-3} = V_{t} - 3\frac{V_{peak+} - V_{t}}{4} \\
& V_{t+1} = V_{t} + \frac{V_{peak+} - V_{t}}{4} \\
& V_{t+2} = V_{t} + 2\frac{V_{peak+} - V_{t}}{4} \\
& V_{t+3} = V_{t} + 3\frac{V_{peak+} - V_{t}}{4} \\
\end{split}
\end{equation}

\begin{table}[h!]
	\caption{The mapping table of 3-bit soft-decision filter}
	\label{table_3}
	\begin{tabular}{ccc}
		\hline
		\bfseries Comparator &\bfseries Range &\bfseries Output LLR values\\ \hline
		0	&[$V_{peak+}$ ;  $V_{t+3}$]     &  1.2017 \\
	 	1	&[$V_{t+3}$   ;  $V_{t+2}$]     &  0.3630 \\
		2	&[$V_{t+2}$   ;  $V_{t+1}$]     &  0.2185 \\ 
		3	&[$V_{t+1}$   ;  $V_{t}$]       &  0.0656 \\	
		4	&[$V_{t}$     ;  $V_{t-1}$]     & -0.0702 \\
		5	&[$V_{t-1}$   ;  $V_{t-2}$]     & -0.2116 \\
		6	&[$V_{t-2}$   ;  $V_{t-3}$]	    & -0.3547 \\
		7	&[$V_{t-3}$   ; $V_{peak-}$]	& -1.1943 \\ \hline
	\end{tabular}
\end{table}

\subsubsection{Successive cancellation (SC) Polar decoder and descrambler}
\label{polardecoder}

Fig.\ref{fig8} shows the proposed hardware architecture of our non-RLL VLC receiver. Firstly, as explained in Section \ref{filter}, the proposed 3-bit soft-decision filter enables applying a soft-decoding Polar decoder at the VLC receiver to increase the reliability of the system. Indeed, the soft-decision filter passes 256 LLR values to the parallel inputs of the SC polar decoder. Compared with conventional architectures of the Polar decoder, our implemented polar decoder (256;158) has three unusual features. Firstly, 8 layers of processing elements (PEs) are purely processed by the combinational logic, because we have removed all intermediate registers to reduce the decoding latency. Secondly, the last stage's PE is modified to output two decoded information bits every each clock cycle. These two modified points have been presented in our previous work \cite{Phuc}. Thirdly, we have implemented a partial sums generator (PSG) based on Polar encoders with various code-length sizes, and integrate the PSG into the scheduling control block of the SC Polar decoder. The decoded data is converted to serial form and is descrambled by a descrambler. Finally, the frame decapsulation block removes the SOF and EOF fields to extract the ID data.

\section{Results and discussion}
\label{results}

\subsection{Flicker mitigation results}
\label{fmresults}

In the previous non-RLL solution work based only on a polar encoder \cite{Fang}, the authors have demonstrated the fluctuation of the code weight distribution around the 50\% dimming level. Specifically, in the case of a polar encoder with 2048-bit codewords, the percentage of one bits was reported to fluctuate in the range of (42.1875\%, 57.8125\%).
\begin{figure*}[h!]
	\centering
	\includegraphics[width=3.5in]{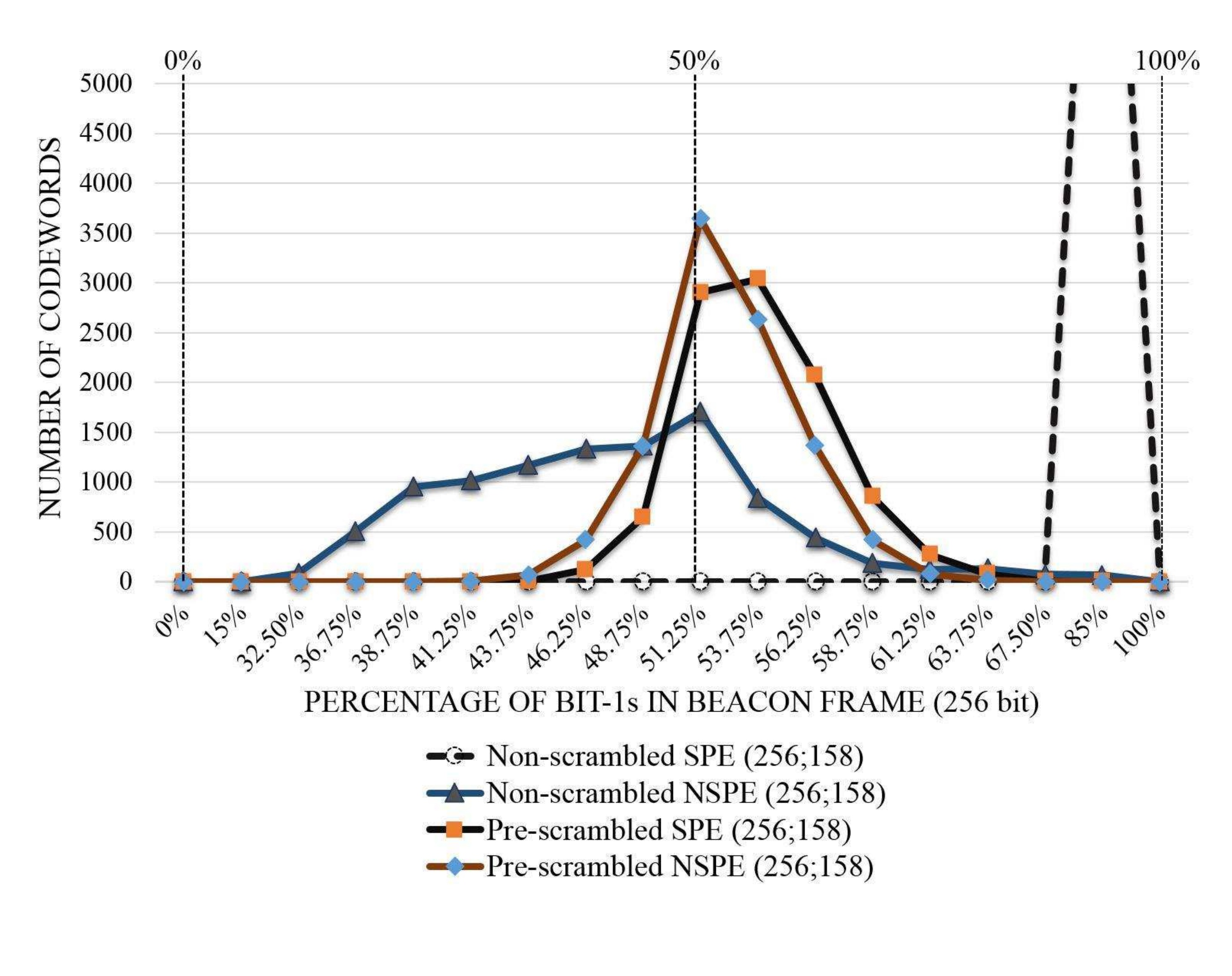}
	\caption{\csentence{}
		Output bit probability distributions of non-scrambled/ pre-scrambled NSPE and SPE}
	\label{fig9}
\end{figure*}

\begin{figure*}[h!]
	\centering
	\includegraphics[width=3.5in]{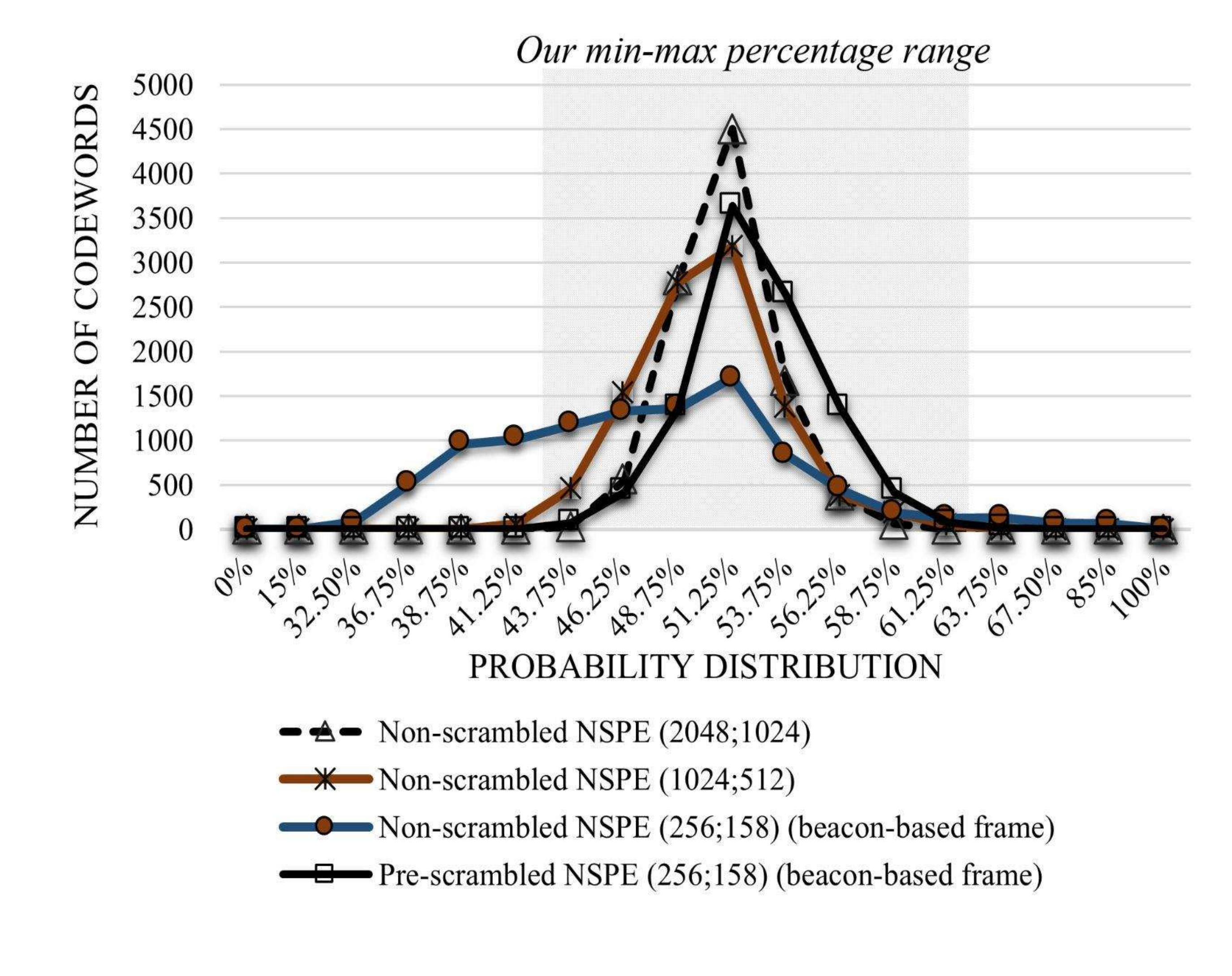}
	\caption{\csentence{}
		Output bit probability distributions of pre-scrambled and non-scrambled NSPEs with long and short codeword lengths}
	\label{fig10}
\end{figure*}

However, we have found that this fluctuation range can only be achieved when the proportions of 1\char`\'s and 0\char`\'s in the input data (before the FEC encoder) are both equal to approximately 50\%. Unfortunately, the bit ratio of the input data is unknown beforehand because of the randomness of these data, and this input bit ratio greatly affects the output bit ratio of the FEC encoder. In this paper, we evaluate our proposed method using a worst-case input bit ratio corresponding to 10\% zero bits and 90\% one bits. A simulation was performed using 10,000 158-bit data frames. If the minimum and maximum bit ratios are included, the real fluctuation range of the output bit probability distribution of a polar encoder with 2048-bit codewords is (41.25\%, 61.25\%). 
From the experimental results presented in Fig.\ref{fig9}, we can also see that the bit distribution of an NSPE shows a little bit more centralized than that of a SPE regardless of whether pre-scrambling is applied. Especially when a pre-scrambler is not used, the probability distribution of the SPE tends toward 85\% one bits because it is greatly affected by the input bit probability distribution. Fig.\ref{fig9} shows the impact of a pre-scrambler on the output bit probability distribution of the NSPE and SPE. Notably, DC balance is not guaranteed in the case of a (256;158) polar code if a pre-scrambler is not applied because the encoder's output bit probability distribution is spread over a large range of percentages (32.5\%, 85\%). However, when a pre-scrambler is used, the output fluctuation range of the pre-scrambled (256;158) polar encoder is (41.25\%, 63.75\%), whereas the fluctuation ranges of polar encoders with codeword lengths of 2048 and 1024 are (41.25\%, 61.25\%) and (38.75\%, 67.5\%), respectively, which is shown in Fig.\ref{fig10}. Thus, pre-scrambling causes the output bit probability distribution of a (256;158) NSPE to be approximately equal to those of (1024;512) and (2048;1024) encoders. 
\begin{figure*}[h!]
	\centering
	\includegraphics[width=3.5in]{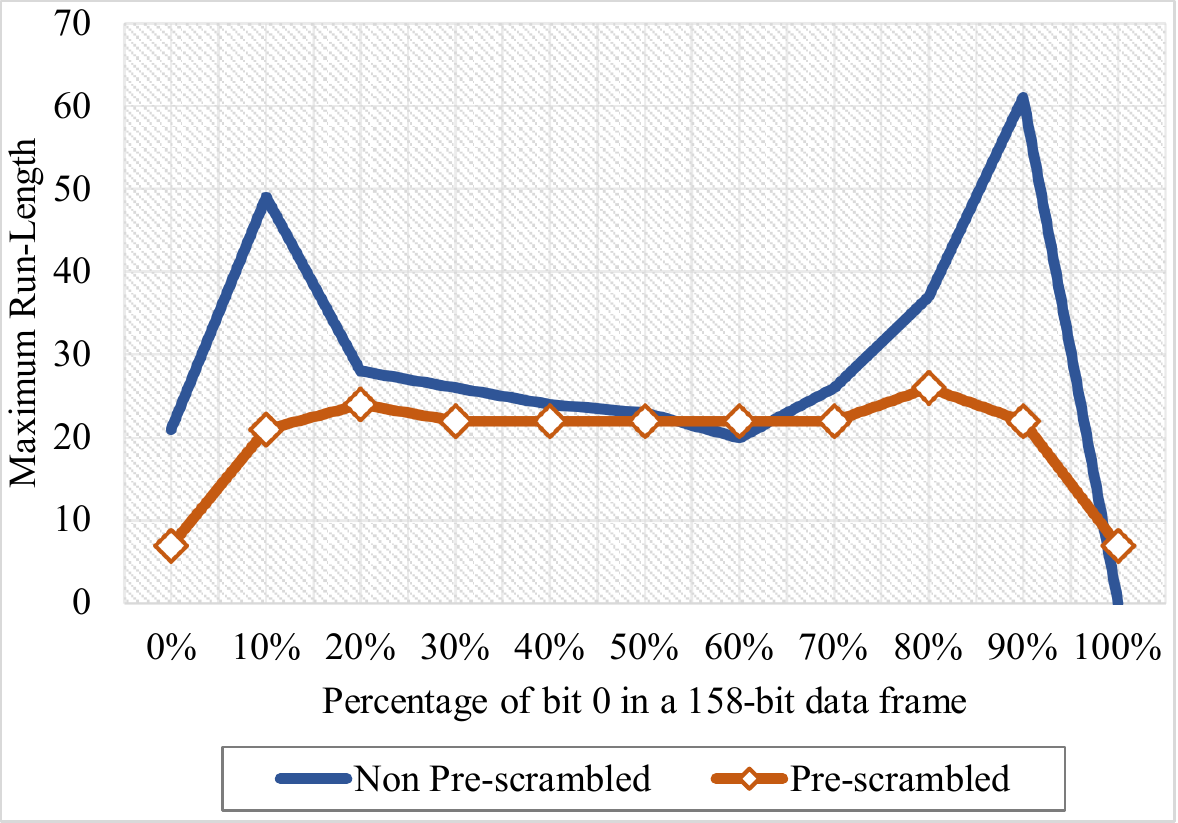}
	\caption{\csentence{}
		Run-length reduction performance of the pre-scrambled NSPE}
	\label{fig11}
\end{figure*}
\begin{figure*}[h!]
	\centering
	\includegraphics[width=3.5in]{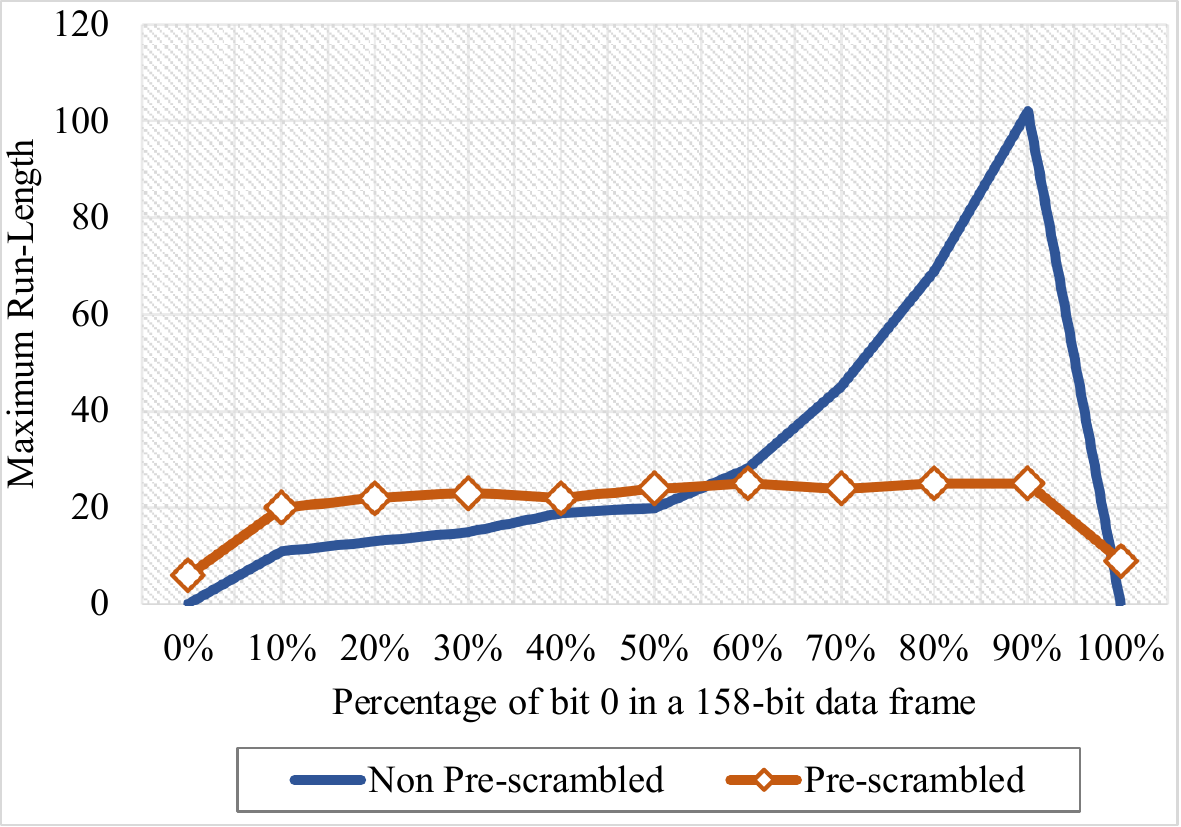}
	\caption{\csentence{}
		Run-length reduction performance of the pre-scrambled SPE}
	\label{fig12}
\end{figure*}

The bit-probability distribution results presented in Fig.\ref{fig9} and Fig.\ref{fig10} demonstrate that a pre-scrambler combined with a Polar encoder is useful for ensuring faster convergence to a centralized output bit probability distribution. Accordingly, DC balance can be guaranteed in VLC-based beacon systems with a short data frame length of 158 bits. Compared with the non-RLL DC-balance solution based only on a polar encoder with 2048-bit codewords presented in \cite{Fang}, the proposed method can achieve the same output bit probability distribution with a codeword length that is shorter by a factor of 8. 

Evaluating the flicker mitigation also requires a consideration on run-length of all packets. In the MATLAB simulation model, we have sent 10000 frames with the percentage of bit-0 in each 158-bit data frame changes from 0\% to 100\%. The results have been presented in graphs at Fig.\ref{fig11} and Fig.\ref{fig12}. Accordingly, the maximum run-lengths are reduced remarkably in case of pre-scrambled technique is applied for Polar encoders. Specifically, the maximum run-length reduction gain that NSPE can achieve is 1.9; while an even better effect that SPE has achieved, is the gain of 4.08 when 90\% of bit-0 appears in a data frame. About the relationship between maximum run-length and minimum flicker mitigation frequency, we introduce a simple equation (Eq.(\ref{eq:minfrequency})) that may be useful to estimate the minimum transmit frequency.       
Specifically, $F_{minFM}$ is the minimum frequency that flicker mitigation is guaranteed; \textit{maxRL} is the maximum run-length and \textit{MFTP} is stated around 5 ms \cite{Fang}. Hence, the minimum frequency that the flicker mitigation in our proposed system is guaranteed is 2.5 Khz, which is still much smaller than the minimum frequency defined in \cite{Sridhar}. 

\begin{equation}\label{eq:minfrequency}
F_{minFM} = \frac{1}{\text{MFTP . maxRL}}
\end{equation}

\subsection{Hardware implementation results}
\label{hwresults}

\begin{table}[h!]
	\caption{FPGA synthesys results of our VLC transmitter and receiver}
	\label{table_4}
	\begin{tabular}{ccc}
		\hline
		\bfseries  &\bfseries Transmitter &\bfseries Receiver \\ \hline
		Device	& Cyclone IV      & Cyclone IV \\
		Model	& 1200mV 0C	    &  1200mV 0C  \\
		Fmax	& 382.85 Mhz     &  29.31 Mhz \\ 
		LE/LUT	& 1896/114480 (1.65\%)       &  12134/114480 (10.6\%) \\	
		Registers	& 1879        & 3109 \\
		Memory bits	& 0     & 1152  \\ \hline
	\end{tabular}
\end{table}

\begin{table}[h!]
	\caption{Resource summary of our VLC transmitter and its function blocks}
	\label{table_5}
	\begin{tabular}{ccccc}
		\hline
		\bfseries Instance &\bfseries Logic Cells &\bfseries Register &\bfseries LUT/Reg.LCs \\ \hline
		Polar encoder	& 1437 (160)   & 1292 (158) & 712(75)   \\
		Frozen Inserter	& 158 (158)	   &  158 (158) & 712 (149) \\
		Parallel to Serial	& 259 (259) &  258 (258) &  193 (193)  \\ 
		Prescrambler	& 5 (5)       &  4(4)        &   5(5)       \\	
		Serial to Parallel	& 178 (178)  & 167 (167) & 168 (168) \\ 
		\bfseries Total &\bfseries 1896 (0) &\bfseries 1879 (0) &\bfseries 946 (0) \\
		\hline
	\end{tabular}
\end{table}

\begin{table*}[h!]
	\caption{Resource summary of our VLC receiver and its function blocks}
	\label{table_6}
	\begin{tabular}{ccccc}
		\hline
		\bfseries Instance &\bfseries Logic Cells &\bfseries Register &\bfseries Mem. bit &\bfseries LUT/Reg.LCs \\ \hline
		Soft-dec. Filter	&   1341 (1341)    &  1301 (1301) &  1152   & 	1300 (1300) \\
		Polar Decoder    	& 	10519 (3192)   &  1545 (1537) &   0 	&	2748 (1522) \\
		Parallel to Serial	&      267 (267)   &  258 (258)	  &   0	    &  	249 (249)	\\ 
		Descrambler	        &      7 (7)       &    5 (5)     &   0     &   5(5)	    \\	 
		\bfseries Total &\bfseries 12134 (0) &\bfseries 3109 (0)&\bfseries 1152  &\bfseries 4302 (0)      \\
		\hline
	\end{tabular}
\end{table*}

The block diagram of the proposed architectures of the VLC transmiter and receiver are presented in Fig.\ref{fig4}, Fig.\ref{fig5}, Fig.\ref{fig6}. The proposed hardware architecture is described by Verilog HDL language before it is synthesized by Quartus II software. The selected targeted device is Cyclone IV EP4CE115F29C7 FPGA. FPGA synthesis results of our VLC transmitter and receiver are given in Table \ref{table_4}. It can be seen that the consumed resource of the receiver is much more than the transmitter one. The consumed LE/LUT of our transmitter takes only 1.65\% of the Cyclone IV FPGA. This result implies that there are around 60 VLC transmitters can be implemented on the same FPGA. Regarding to this result, we expect that multi-VLC-transmitters could be implemented on one FPGA to reduce the cost and enhance the performance of the VLC system (Section \ref{whyHW}).    

Table \ref{table_5},\ref{table_6} summarize the logic resource that components of our transmitter and receiver have consumed. At the transmitter, the Polar encoder is the biggest block which takes around 76\% of the whole VLC transmitter. On the other hand, the prescrambler only takes less than 1\% resource of the whole transmitter but it provides an effective solution in centralizing bit-probability distribution (Section \ref{fmresults}). At the receiver, the Polar decoder block is the most heavy one. Specifically, it takes more than 85\% resource of our receiver. Whereas, the soft-decision filter occupies around 11\%, and the descrambler only takes an unnoticeable amount of logic cells. The maximum frequency of the VLC transmitter can be up to more than 382 Mhz because the polar encoder is created mostly from modulo-2 computations which are simple elements. On the contrary, the receiver's maximum frequency is reported at 29.31 Mhz. The reason for this relatively low frequency is as we mentioned in Section \ref{polardecoder}; due to the nature of VLC-based beacon systems, high-throughput is not a high-priority criteria in VLSI architecture design. Therefore, we have implemented a architecture based on combinational logic for the network of PEs; hence, this causes an increase in delay of PEs-network and a deduction of frequency. Moreover, the synthesis results of consumed LEs/LUTs of transmitter and receiver are strongly affected by the code-length of Polar encoder/decoder. Specifically, We have noticed that the amount of consumed LEs/LUTs almost duplicate when the code-length increase 2 times. Table \ref{table_7} introduces some ASIC synthesis results of our VLC transmitter and receiver. Specifically, we have utilized the Synopsys' Design Compiler RTL synthesis tool in which the VDEC's Rohm technology library 180nm is selected. Besides results of power consumption and area are reported, we also evaluate the throughput, energy-per-bit and hardware efficiency of implemented transmitter/receiver based on equations given in this paper's appendix (\ref{eq:Energyperbit}, \ref{eq:Energy}, \ref{eq:Hardware}) \cite{dizdar2016high}.   

\begin{table}[h!]
	\caption{ASIC synthesis results of our VLC transmitter and receiver}
	\label{table_7}
	\begin{tabular}{ccc}
		\hline
		\bfseries  &\bfseries   Transmitter      &\bfseries Receiver \\ \hline
		Technology [$nm$]		& 180          & 180 \\
		Voltage    [$V$]		& 1.8	       &  1.8  \\
		Area      [$\mu m^2$]	& 48761.39     &  573724.56 \\ 
		Frequency [$Mhz$]       &  25 	 	   & 25		\\	
		Power     [$mW$]     	& 1.3137       & 3.5022 \\
		Throughput [$Mb/s$]		& 15.38        & 16.58 	\\ 
		Energy-per-bit [$pJ/b$] & 85.42        &  211.2    \\ 
		Hardware Efficiency [$Mb/s/mm^2$]      & 315.41     & 28.75  \\ 
		Latency [$clocks$]      & 160           & 386  \\ \hline
	\end{tabular}
\end{table}


\begin{figure*}[h!]
	\centering
	\includegraphics[width=4in]{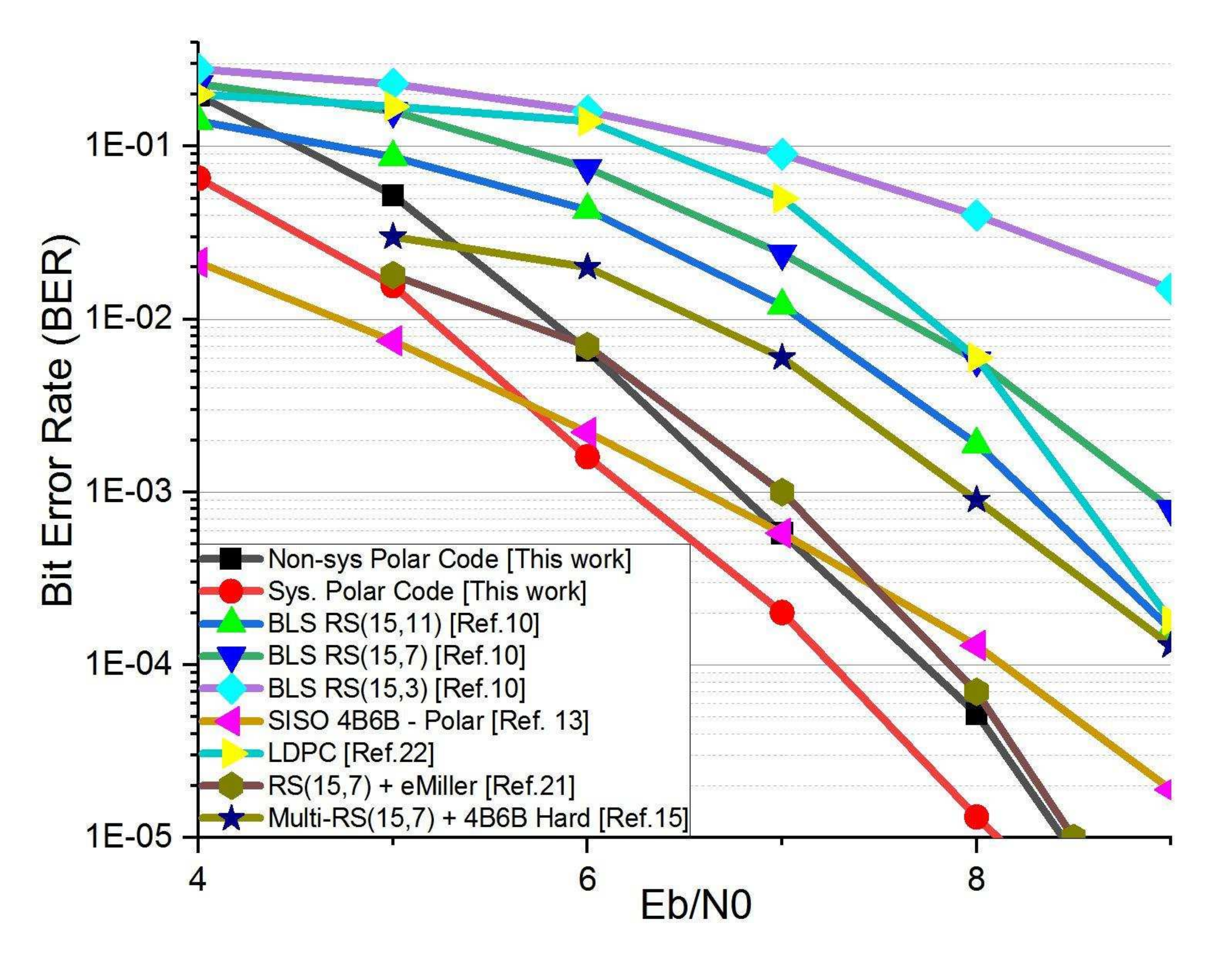}
	\caption{\csentence{}
		BER performances of the proposed VLC system with some comparisons in real VLC-AWGN channel}
	\label{fig13}
\end{figure*}

\subsection{Bit error rate (BER) and frame error rate (FER) performances}

\begin{figure*}[h!]
	\centering
	\includegraphics[width=4in]{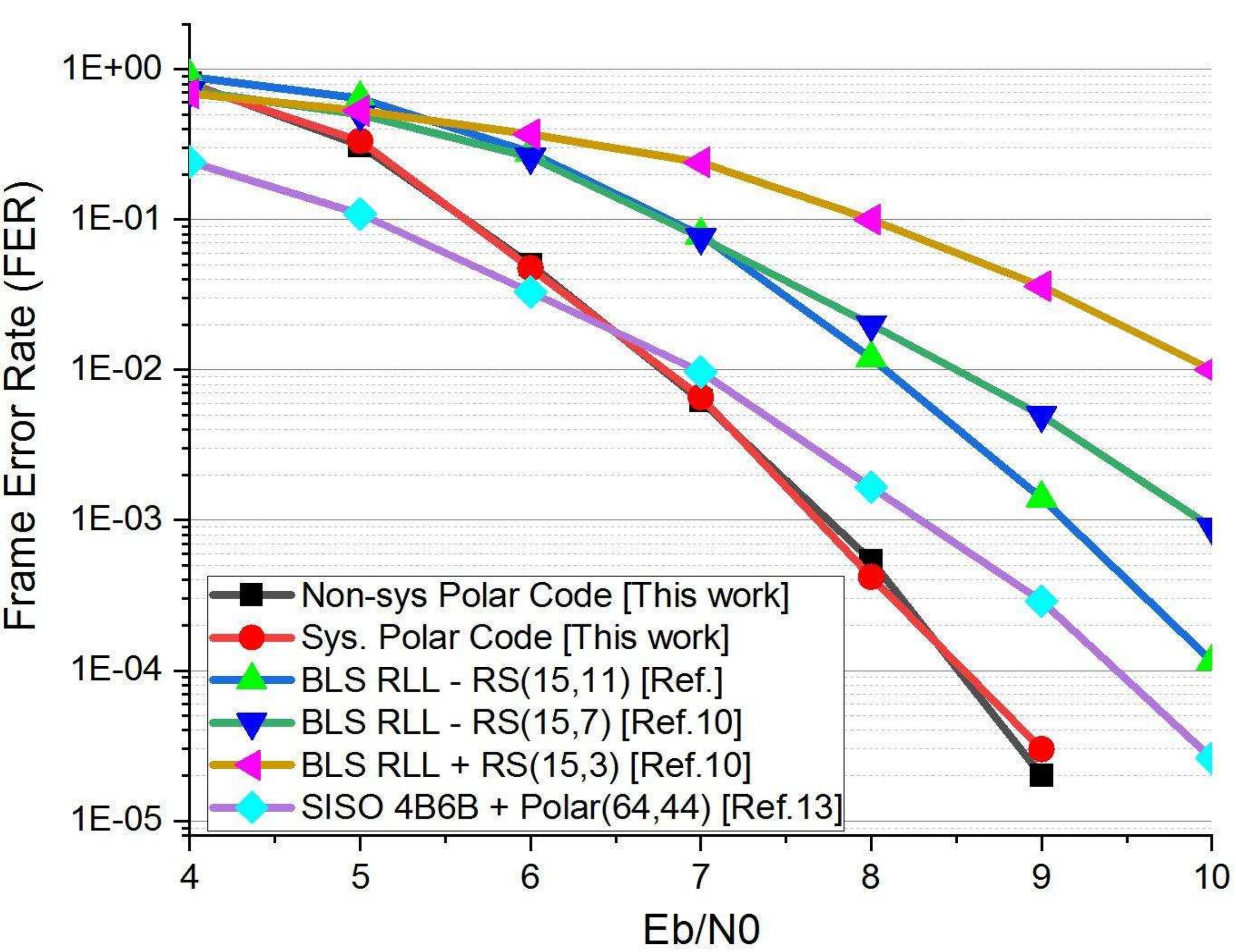}
	\caption{\csentence{}
		FER performances of the proposed VLC system and some related works in real VLC-AWGN channel}
	\label{fig14}
\end{figure*}

Fig.\ref{fig13} shows the BER performances of our proposed VLC system in which both systematic and non-systematic Polar codes are applied to evaluate the BER performances. We have selected some typical joint FEC-RLL and non-RLL FEC solutions for comparison. It can be seen that the BER performances of our proposed system outperform current related works. Specifically, Fig.\ref{fig13} shows that at a code-rate = 0.62, our non-RLL Polar-code-based system outperforms RS-code-based ones at code-rates 0.49 (11/15 * 4/6), 0.31 (7/15 * 4/6) and 0.13 (15/3 * 4/6) which are mentioned in \cite{Wang2,Le}. Also, in Fig.\ref{fig13}, we also put BER performances of other related works mentioned in \cite{Le,Wang1,Xuanxuan,Sunghwan1} into the same graph with our BER performance lines. It can also be noticed that, our prescrambled non-RLL Polar code-based solutions have preeminent BER performances although their humble code-rates compared with the related works. Furthermore, an evaluation of frame error rate (FER) has been presented in Fig.\ref{fig14} in which our non-RLL solutions also surpass related works introduced in \cite{Wang2,Le}. However, although systematic Polar decoder always shows a better BER performance than the non-systematic decoder does; the FER performance of these two decoders are always equal in all cases. Actually, this is not a strange discovery cause it has been mention in previous systematic Polar decoder work \cite{Arikan}.

\section{Conclusions}
We have introduced a non-RLL flicker mitigation solution which consists of a pre-scrambler based on a simple generating polynomial combined with a Polar encoder. The proposed method has a centralized bit probability distribution with the distribution range is determined in (43.75\% - 63.75\%). Moreover, the maximum run-length is reduced up to 4.08 times when pre-scrambler is applied with a SPE; and up to 1.9 times when it is applied with a NSPE. Therefore, DC-balance can be maintained even with the short data frames used in VLC-based beacon systems. Moreover, the non-RLL nature of the proposal reduces the complexity of both VLC transmitter/receiver with great improvements on information code-rate. Besides, we have introduced a soft-decision filter which can help the soft-decoding of polar code is implemented in real VLC receiver prototypes to enhance the error-correction performance. As a result, BER and FER performances of the proposed system have outperformed current approaches while remaining a good code-rate (0.62). In addition, we have introduced a couple of hardware architectures for the proposed non-RLL VLC transmitter and receiver which their FPGA and ASIC synthesis results are given in details.

\appendix
\section{}

\begin{equation} \label{eq:Energyperbit}
\text{Thoughput [b/s]} = \dfrac{\text{$N$ [b]}}{\textrm{$D_N$ [sec]}} 
\end{equation}

\begin{equation} \label{eq:Energy}
\text{Energy-per-bit [J/b]} = \dfrac{\text{Power [W]}}{\text{Thoughput [b/s]}} 
\end{equation}

\begin{equation} \label{eq:Hardware}
\text{Hardware Efficiency [b/s/$m^2$]} = \dfrac{\text{Thoughput [b/s]}}{\text{Area [$m^2$]}}
\end{equation}

\begin{backmatter}
\section*{Abbreviations}
VLSI: Very Large Scale Integration; RLL: Run-length limited; VLC: Visible Light Communication; IPS: Indoor Positioning System; LED:    
Light-emitting Diode; MCU: Microcontroller; SoC: System on Chip; LLR: Log-likelihood Ratio; FEC: Forward Error Correction; BER: Bit Error Rate; FER: Frame Error Rate; FPGA: Field-programmable Gate Array; ASIC: Application-specific Integrated Circuit; VPPM: Variable Pulse Position Modulation; OOK: On-off Keying; OFDM: Orthogonal Frequency Division Multiplexing; TX: Transmit; RX: Receive; GPS: Global Positioning System; LBS: Location-based Service; UWB: Ultra-wideband; RFID: Radio-Frequency Identification; RF: Radio Frequency; ID: Identity, Identification; JEITA: Japan Electronics and Information Technology Industries Association; SOF: Start of frame; EOF: End of frame; DC: Direct current; MFTP: Maximum flickering time period; GPIO: General Purpose Input/Output; RS: Reed-Solomon; URC: Unity-Rate Code; IRCC: IRregular Convolutional Code; RM: Reed-Muller; LDPC: Low-density parity check; CS: Compensation symbol; SPE: Systematic Polar Encoder; NSPE: Non-systematic Polar Encoder; SC: Successive Cancellation; S2P: Serial to Parallel; P2S: Parallel to Serial; AWGN: Additive White Gaussian Noise; ADC: Analog-to-digital Converter; PE: Processing element; PSG: Partial Sums Generator; HDL: Hardware Description Language; LE: Logic Element; LUT: Lookup Table; RTL: Register-transfer level; VDEC: VLSI Design and Education Center.         	
\section*{Competing interests}
  The authors declare that they have no competing interests.

\section*{Author's contributions}
All authors contributed to the work. THT and YN are the supervisors of this project. HTH is collaboration member of this project. DPN conceived and designed the study. DPN carried out most of the analyses. DPN drafted the manuscript. DDL performed some experiments of BER,PER and soft-decision filter work. All authors have read and approved the manuscript. 

\section*{Funding}
  This work  was supported by JSPS KAKENHI Grant Number JP16K18105 and NAIST's Global Cooperation Research Project Fund. 

\bibliographystyle{bmc-mathphys} 
\bibliography{bmc_article}      

\end{backmatter}
\end{document}